\documentclass{aipproc}

\usepackage{graphicx}

\newcommand{\onepion}{single-pion\ }

\newcommand{\GeV}{\; \mathrm{GeV}}

\newcommand{\dd}{\mathrm{d}}

\layoutstyle{8x11double}

\begin{document}

\title{Neutrino-nucleus interactions}

\author{U. Mosel}{address={Institut f\"ur Theoretische Physik, Universit\"at Giessen, Germany}}
\author{O. Lalakulich}{address={Institut f\"ur Theoretische Physik, Universit\"at Giessen, Germany}}

\begin{abstract}
Interactions of neutrinos with nuclei in the energy ranges relevant for the MiniBooNE, T2K, NO$\nu$A, MINER$\nu$A and MINOS experiments are discussed. It is stressed that any theoretical treatment must involve all the relevant reaction mechanisms: quasielastic scattering, pion production and DIS. In addition, also many-body interactions play a role. In this talk we show how a misidentification of the reaction mechanism can affect the energy reconstruction.  We also discuss how the newly measured pion production cross sections, as reported recently by the MiniBooNE collaboration, can be related to the old cross sections obtained on elementary targets. The MiniBooNE data seem to be compatible only with the old BNL data. Even then crucial features of the nucleon-pion-Delta interaction are missing in the experimental pion kinetic energy spectra. We also discuss the meson production processes at the higher energies of the NO$\nu$A, MINER$\nu$A and MINOS experiments. Here final state interactions make it impossible to gain knowledge about the elementary reaction amplitudes. Finally, we briefly explore the problems due to inaccuracies in the energy reconstruction that LBL experiments face in their extraction of oscillation parameters.
\end{abstract}

\maketitle

\section{Introduction}
Investigations of neutrino interactions with nucleons can on one hand give valuable information on the axial properties of the nucleon. While the vector couplings of nucleons and their resonances have been explored since many decades in high-precision electron scattering experiments, the axial properties of nucleons are still connected with much larger uncertainties. On the other hand, such experiments also give novel information on neutrino properties, i.e.\ their masses and mixing angles. For both types of analyses it is mandatory to know the neutrino energy. This is not an easy task since neutrinos are produced as secondary decay products of primarily produced hadrons and thus the neutrino beams are not monoenergetic. This requires a reconstruction of the neutrino energy from final state observations. Such reconstructions have to rely on the assumption of a well-defined reaction mechanism.

The identification of this mechanism becomes difficult when the targets are not nucleons, but nuclei. Apart from the ever-present Fermi-smearing and the action of the Pauli-principle there can be significant final state interactions (FSI) that may hide the special characteristics of the primary interaction vertex. In addition, experiments usually do not observe the full event. An example is given by Cherenkov counter experiments which can only see pions and high-energy protons. Indeed, in the MiniBooNE (MB) experiment all events with 1 $\mu$ and 0 pions were identified as quasielastic-like scattering of a neutrino on a quasifree nucleon and the energy was reconstructed on that basis.

However, the actual reaction mechanism may be significantly more complicated. There can be events in which first a pion was produced which then got stuck in the nucleus due to FSI and thus did not get out to the detector. Such an event contributes to the quasielastic-like event rate \cite{Leitner:2010kp}. In addition, there can be primary interactions of the neutrino with 2 nucleons (so-called 2p-2h interactions) \cite{Martini:2009uj}. All these latter events were identified as QE scattering events by the MB experiment and thus had to be subtracted out from the data by means of event generators.

\begin{figure}[b!ht]
\includegraphics[width=\columnwidth]{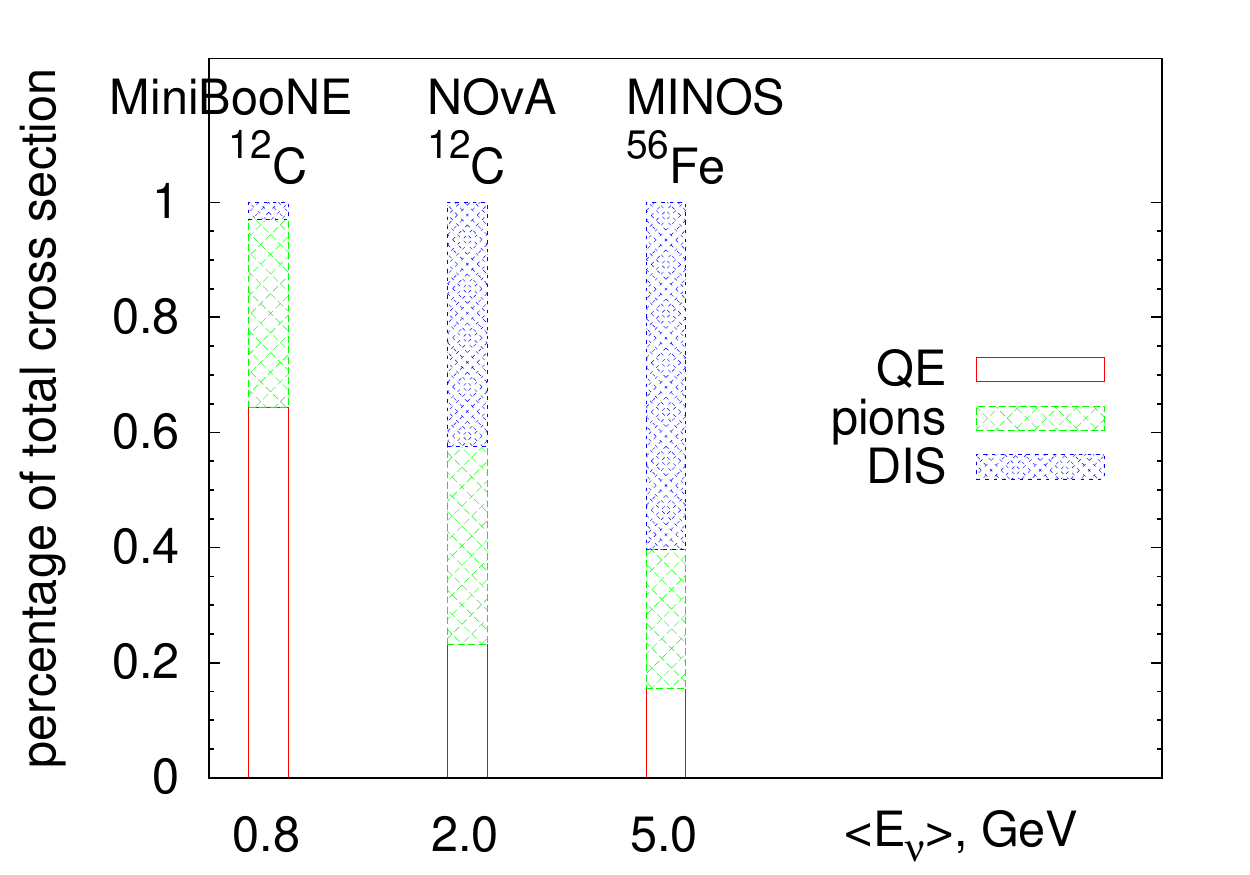}
\caption{(Color online) Reaction mechanisms at the 3 experiments presently running at Fermilab. The numbers below each of the vertical bars give the average neutrino energy for that experiment. The contribution labeled 'pions' contains all resonance as well as background contributions.} \label{fig:compare-origin-MINOS-MB-3}
\end{figure}

These event generators have to be reliable for very different reactions types. This is illustrated in Fig.\ \ref{fig:compare-origin-MINOS-MB-3} which shows that the MB experiment had to deal with QE scattering and pion production, while deep inelastic scattering (DIS) events contributed very little. The situation has changed when one goes to the MINOS experiment; here,  at a mean energy of 5 GeV, quasielastic (QE) scattering, pion production and DIS all contribute approximately with equal size. Any event generator thus has to be able to describe all these different reaction types with an equally good accuracy.

In the following we discuss some aspects of a theoretical treatment of neutrino-induced reactions on nuclei. We start off with a discussion of event-identification and energy reconstruction in connection with QE scattering, then investigate pion production and the special difficulties that appear in the understanding of the present-day experiments MINOS and NO$\nu$A. Finally, we give a brief discussion on how our understanding of reaction mechanisms affects the extraction of oscillation parameters.

All results that we are going to discuss are based on the GiBUU transport model. GiBUU stands out from all the other neutrino event generators in that it aims to solve the Kadanoff-Baym transport equations \cite{Kad-Baym:1962} in the Botermans-Malfliet approximation \cite{Botermans:1990qi} for off-shell transport. It has been widely tested on very different classes of nuclear reactions, starting from relativistic heavy-ion collisions to electron and neutrino-induced reactions. For all details of this model, its theoretical foundation and its practical implementation, we refer to a recent review \cite{Buss:2011mx}. The results to be discussed in this conference report have recently been published in Refs.\ \cite{Lalakulich:2012ac,Lalakulich:2012gm,Lalakulich:2012hs,Lalakulich:2012cj} where details about the calculations can be found.

\section{Event identification}
\begin{figure}[!hbt]
\includegraphics[width=\columnwidth]{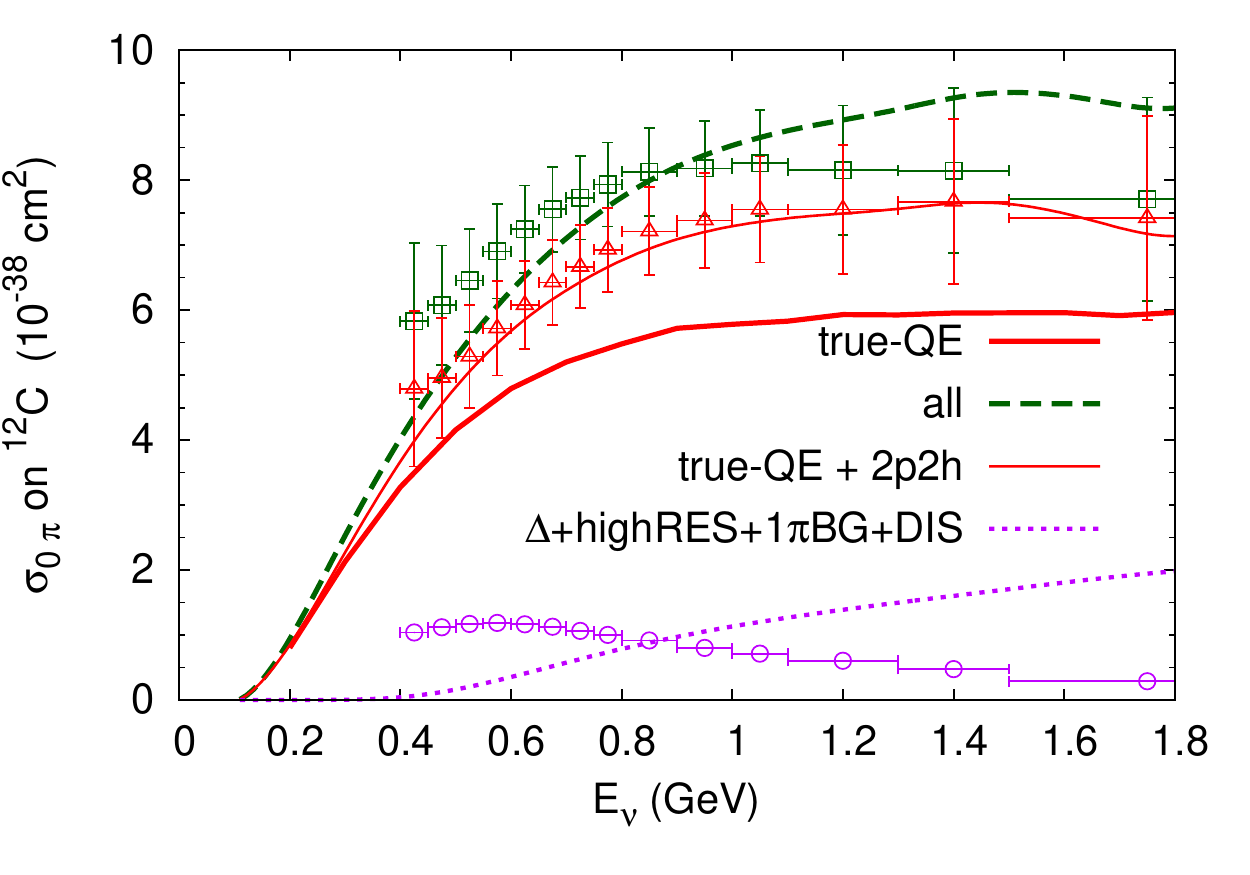}
\caption{(Color online)
QE-like cross section originating from QE and 2p-2h processes only (solid line) and from all processes (dashed line) within
the GiBUU calculations. The thick solid curve gives the true QE cross section. ``Measured`` (sqares) and ''extracted'' (triangles)  MiniBooNE data points are taken from \protect\cite{AguilarArevalo:2010cx}. The difference between them (open circles) is to be compared with the GiBUU 'stuck pion' cross section (dotted line).
All data are plotted vs reconstructed energy, whereas the theoretical curves are plotted vs true neutrino energy. }
\label{fig:all-versusTrue}
\end{figure}

Fig.\ \ref{fig:all-versusTrue} shows both the QE-like and the extracted QE cross sections as obtained in the MB experiment. While the uppermost data points give the actually measured cross section for all events without any pions, dubbed 'QE-like', the lower, extracted points are obtained after subtracting the so-called stuck-pion events, i.e.\ events in which pions or $\Delta$s were first produced, but then got reabsorbed so that they are no longer present in the final state. This subtraction is model dependent, in the case of the MB experiment it has been performed with the NUANCE event generator. The calculated true QE cross section (thick solid line) lies well below even these extracted data. This latter difference has been explained by the presence of 2p-2h excitations \cite{Martini:2009uj,Nieves:2011yp} that in the MB experiment cannot be distinguished from true QE events, because the nucleons are not observed. It has also been shown that the misidentification of the reaction mechanism leads to errors in the reconstructed neutrino energy \cite{Martini:2012fa,Nieves:2012yz,Lalakulich:2012hs}.

\begin{figure}[!hbt]
\includegraphics[width=\columnwidth]{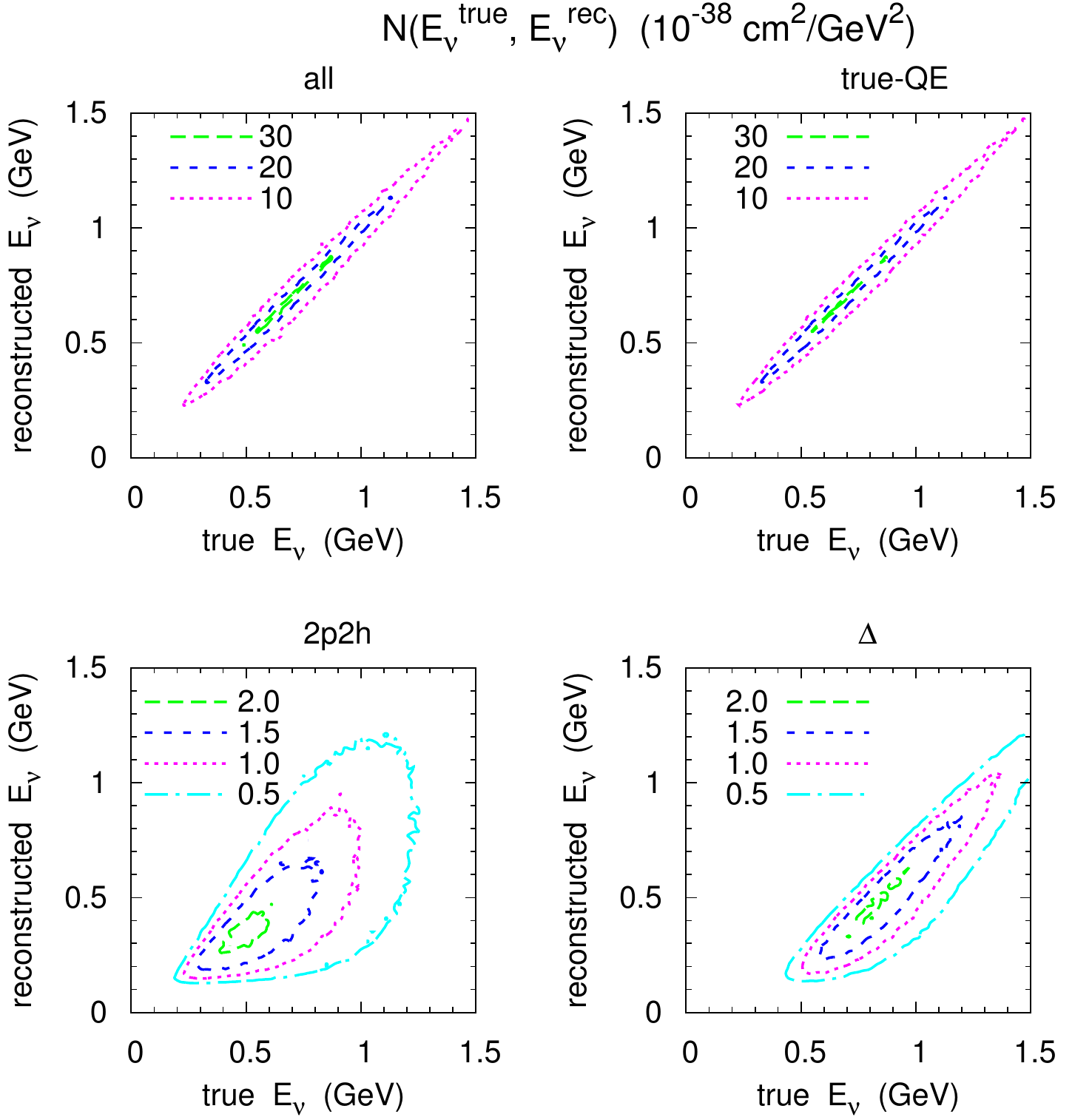}
\caption{(Color online) 2-D density of the QE-like cross section $N(E^{\rm true},E^{\rm rec})$ (the so-called migration matrix)
versus true and reconstructed neutrino energies for all events and for events of various origins, all for the MB flux. Figure taken from \cite{Lalakulich:2012hs}.}
\label{fig:events-reconstruction3D}
\end{figure}

Since the extracted QE data contain a generator dependence due to the background subtraction of stuck-pion events, any analysis of the actual data must involve both QE, pion production and pion absorption. While Ref.\ \cite{Nieves:2012yz} contained a sophisticated treatment of QE only, the work of Ref.\ \cite{Martini:2012fa} did treat pion production in addition, however, without any pion FSI. We have, therefore, performed in \cite{Lalakulich:2012hs} a complete study of all three necessary ingredients. That this is necessary can be seen by going back to Fig.\ \ref{fig:all-versusTrue} which exhibits an astonishing behavior already in the QE-like data. The QE-like and the extracted QE cross sections show most of their difference at the lowest energies whereas they become much closer at the highest energies; this difference is shown by the open circles. As discussed above, this difference should be due to stuck-pion events and the pion production probability steeply increases with beam energy. Thus, the opposite behavior was to be expected. In \cite{Lalakulich:2012hs} we have shown that this unexpected behavior is caused by errors made in the energy reconstruction due to the misidentification of the reaction process. Fig.\ \ref{fig:events-reconstruction3D} shows that for true QE events the reconstruction procedure works very well, but for all other reactions it leads to a lower reconstructed energy than the true energy.
At the same time the functional shape of the stuck-pion cross sections as a function of reconstructed energy is quite different from that of that same cross section as a function of true energy. This is illustrated by the dotted line in Fig.\ \ref{fig:all-versusTrue} which gives the cross section for stuck-pion events as a function of true energy; it indeed increases with energy as it should whereas the same quantity as a function of reconstructed energy starts up high and then decreases as a function of energy (open circle symbols); for a more detailed discussion see \cite{Lalakulich:2012hs}.

\begin{figure}[h!bt]
\includegraphics[width=\columnwidth]{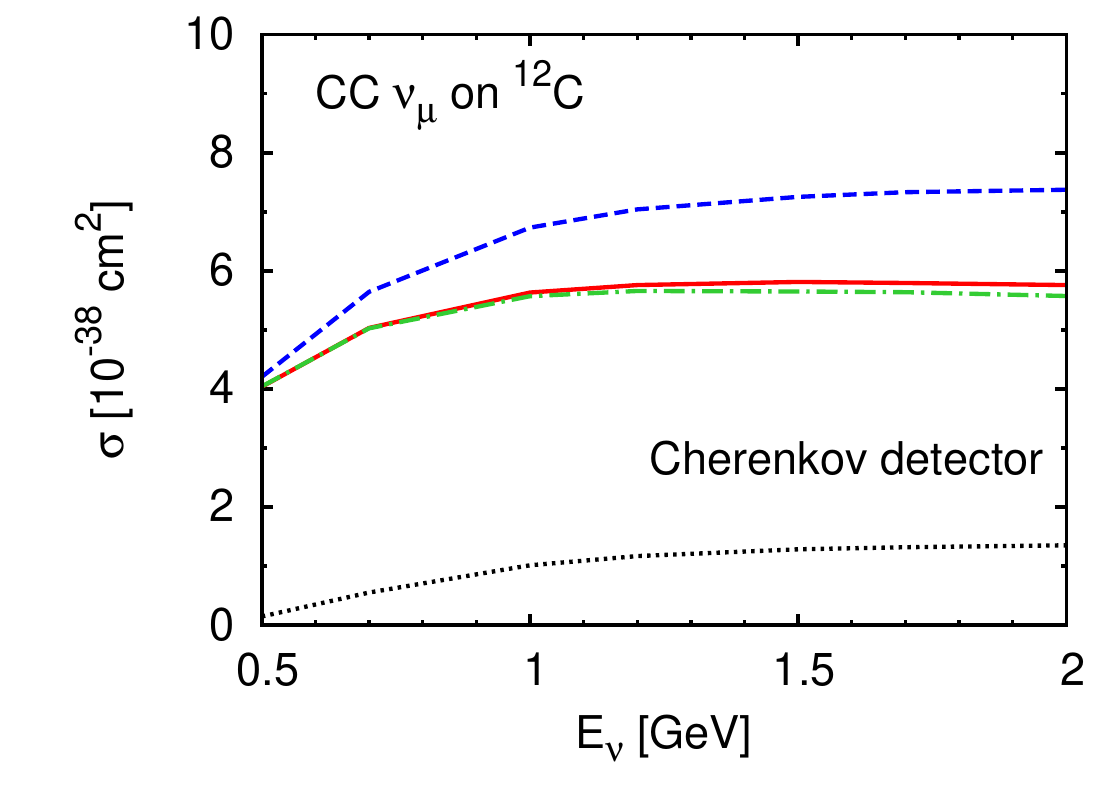}
\caption{(Color online) Total QE cross section on $^{12}C$ (solid
lines) compared to the method used to identify
CCQE-like events in experiments (dashed line). The figure shows the method commonly applied in Cherenkov
detectors.  The contributions to the CCQE-like
events are also classified [CCQE-like from initial QE (dash-
dotted) and from initial $\Delta$ (dotted lines)].  Figure taken from \protect\cite{Leitner:2010kp}.}
\label{fig:Cherenkov}
\end{figure}

\begin{figure}
\includegraphics[width=\columnwidth]{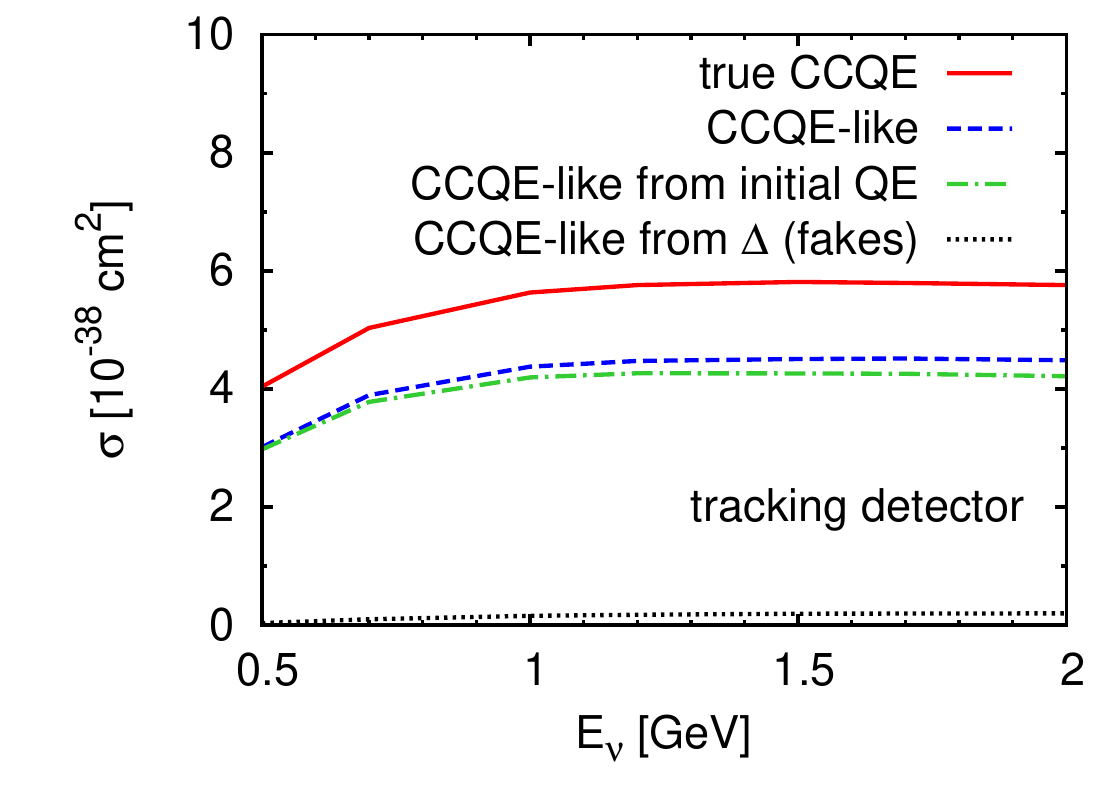}
\caption{Same as Fig.\ \ref{fig:Cherenkov}, but for Tracking Detector}
\label{fig:Tracking}
\end{figure}

In view of these rather large uncertainties in the extraction of physical properties one is tempted to look for better strategies to identify a certain reaction type. In particular, for the case of QE scattering  it is worthwhile to remember that tracking detectors are much better suited to correctly identify a  reaction mechanism because they rely not only on the absence of a pion, but also on the presence of one outgoing nucleon. This is illustrated in Figs.\ \ref{fig:Cherenkov},\ref{fig:Tracking}.
The figures shows nicely that Cherenkov detectors always measure a too-high cross section, because other event types cannot be distinguished from true QE. Tracking detectors miss part of the total cross section because they will not see events in which the initial proton kicks out a second proton or undergoes charge exchange into a neutron. However, tracking detectors allow a much cleaner event identification than Cherenkov counters: nearly all the events ascribed to QE scattering are indeed from that source.

\begin{figure}[!hbt]
\includegraphics[width=0.7\columnwidth]{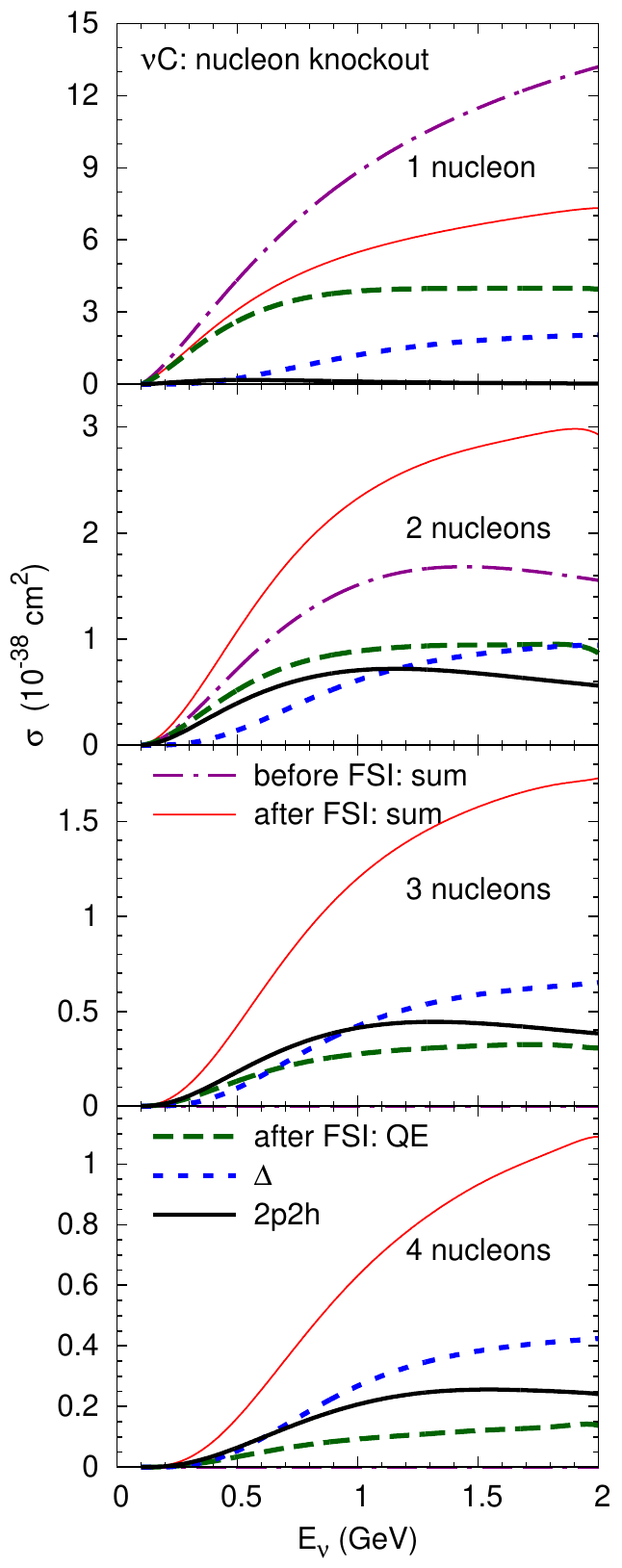}
\caption{(color online) Cross-section for multi-nucleon knock-out as a function of neutrino energy.
In each case the total cross section is given by the thin solid line.
The QE contribution is indicated by the long-dashed line. The short dashed line shows the $\Delta$ contribution, while the 2p-2h contribution is indicated by the thick solid line. All these lines show the results after FSI, while the dot-dashed curves shows the total cross section before FSI. Figure taken from \protect\cite{Lalakulich:2012ac}.}
\label{fig:sigma(Enu)_multinucleon}
\end{figure}
This is found to be true also when there are 2p-2h processes in the first, primary interaction of the neutrino with the nucleus. The naive expectation that then the knock-out of 2 particles is favored \cite{Sobczyk:2012ms} is overshadowed by the strong FSI effects on the outgoing nucleons. These FSI lead to a sort of 'avalanching' so that the energy of a single nucleon is quickly distributed over many of them. Thus the multiplicity of the outgoing nucleons is only a rather weak indicator of the primary process. This can be seen in the results shown in Fig.\ \ref{fig:sigma(Enu)_multinucleon}. The two-nucleon knock-out contains nearly equal contributions from true QE and initial 2p-2h processes so that the identification of either one will be very difficult, if not impossible. However, the same figure also shows that the one-and-only-one nucleon knock-out channel is -- apart from $\Delta$ contributions -- dominated by true QE.

This channel could thus be used to identify true QE scattering if the $\Delta$ contribution can be controlled.

\section{Pion production around 1 GeV}
The difference between the measured QE-like and the extracted QE cross sections determined by MB is due to stuck-pion events. There is then obviously a close connection between this difference and the pion production cross sections also measured by MiniBooNE \cite{AguilarArevalo:2010bm,AguilarArevalo:2010xt}. Until recently, the only comparisons of theoretical calculations with these data were those in Refs.\ \cite{Lalakulich:2011ne} that both showed a significant underestimate of the experimental data and a significantly different spectral shape of the produced pions. These calculations used the old ANL data as input and the full MiniBooNE flux for both charge states of the pion.
\begin{figure}[h!b]
\includegraphics[width=\columnwidth]{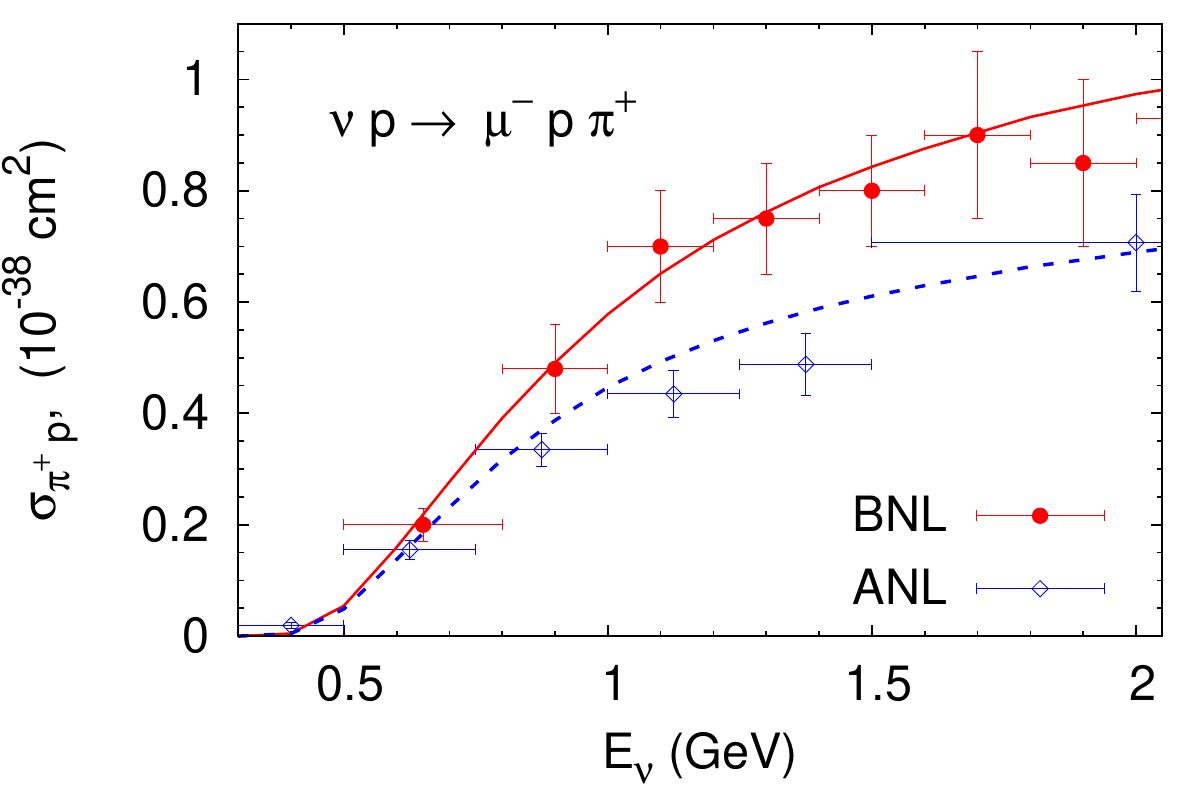}
\caption{(Color online) Single-pion production cross section on a proton
target obtained in the BNL \protect\cite{Kitagaki:1986ct} (circles; solid curve) and the ANL experiments \protect\cite{Radecky:1981fn}  (diamonds; dashed curve) for $1\pi^+$ production. The curves give the lower (ANL-tuned) and upper (BNL-tuned) boundaries on the elementary input as used in GiBUU.}
\label{fig:ANLBNL1pidata+}
\end{figure}

\begin{figure}[h!tb]
\includegraphics[width=\columnwidth]{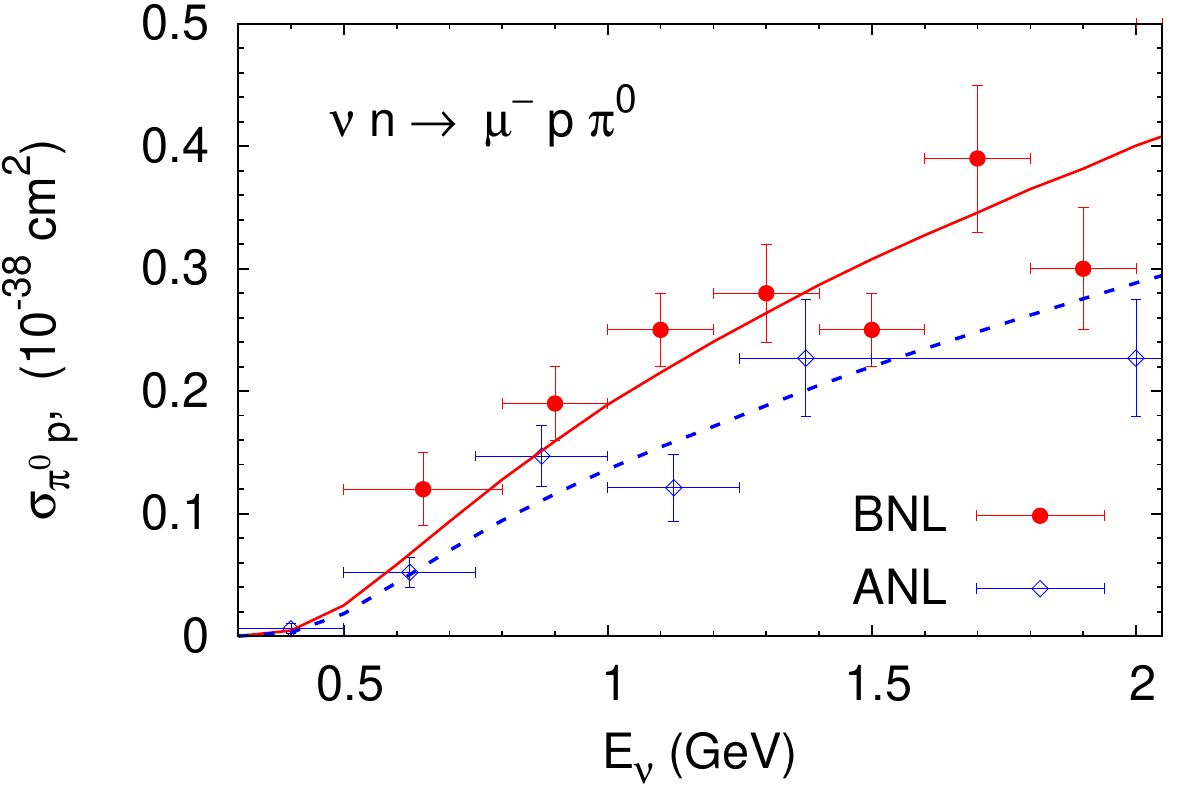}
\caption{Same as Fig.\ \ref{fig:ANLBNL1pidata+} for $1\pi^0$ production.}
\label{fig:ANLBNL1pidata0}
\end{figure}
We have recently extended these studies by using both the ANL and the BNL data \cite{Radecky:1981fn,Kitagaki:1986ct} as an input to the calculations \cite{Lalakulich:2012cj}. Since the BNL data are consistently higher than the ANL data (see Figs.\ \ref{fig:ANLBNL1pidata+},\ref{fig:ANLBNL1pidata0}) we obtain a band of predictions for the pion production cross section. In addition, the calculations for $\pi^0$ production were now redone with the same cut on the neutrino energy $0.5 \GeV < E_\nu < 2.0 \GeV$ as that used in the experimental analysis.

\begin{figure}[!hbt]
\includegraphics[width=\columnwidth]{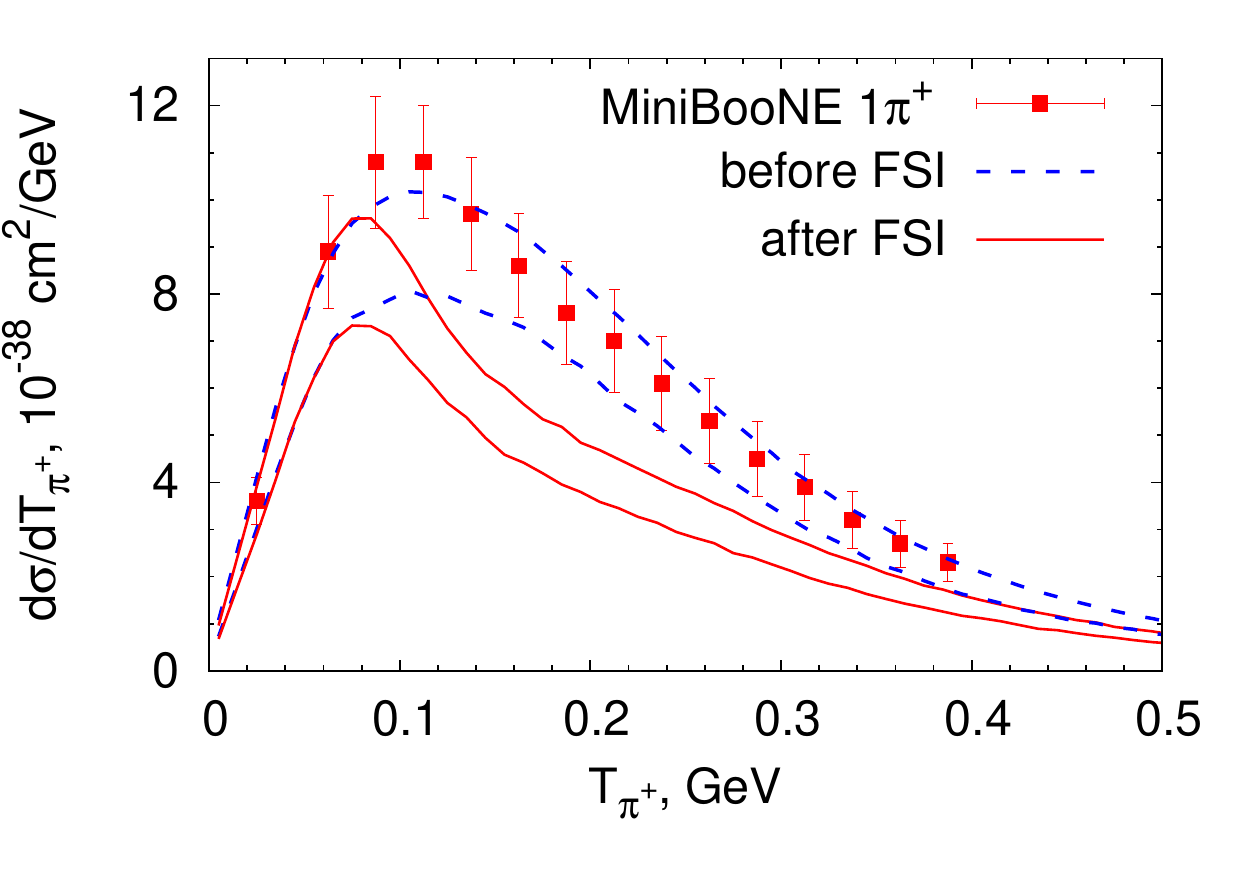}
\caption{Kinetic energy distribution of the outgoing $\pi^+$ for \onepion production at MiniBooNE.
Data are from~\protect\cite{AguilarArevalo:2010bm}. The upper and lower dashed curves give the results before FSI using the BNL and the ANL elementary cross sections, resp. The solid lines give the corresponding results after FSI. Figure taken from \protect\cite{Lalakulich:2012cj}.}
\label{fig:MB-pion-dTkin}
\end{figure}

The results of these calculations are shown in Fig.\ \ref{fig:MB-pion-dTkin}. The difference between the two dashed curves (before FSI) and the two solid curves (after FSI) gives a band of uncertainty due to the uncertainty in the elementary input. While the results before FSI follow the data fairly well, the results after FSI underestimate it significantly and contain a structure in the momentum distribution that is not there in the data. This is surprising since this structure is due to pion absorption through the $\Delta$ resonance and has been experimentally observed in photoproduction of $\pi^0$ mesons on nuclei \cite{Krusche:2004uw}.
\begin{figure}
\includegraphics[width=\columnwidth]{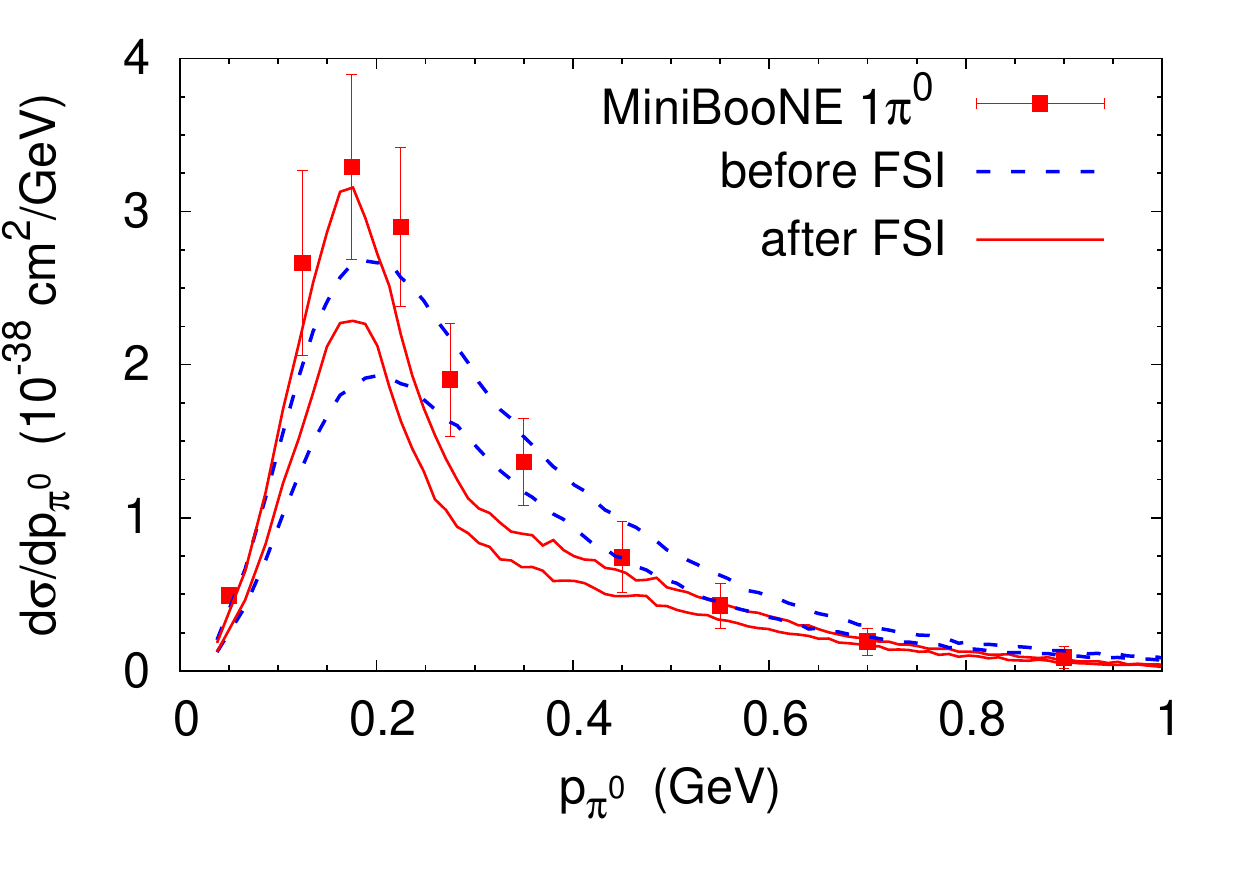}
\caption{Same as Fig.\ \ref{fig:MB-pion-dTkin}, but for the momentum distribution of the outgoing $\pi^0$. Data are from \protect\cite{AguilarArevalo:2010xt}. Figure taken from \cite{Lalakulich:2012cj}.}
\label{fig:MB-pion-dp}
\end{figure}

The MiniBooNE experiment had determined its neutrino flux by using hadronic production cross sections. The analysis of QE scattering by Nieves et al.\ \cite{Nieves:2011yp} has shown that a good description of the MB QE data can be reached when the actual flux is assumed to be 10\% higher than determined by the MB experiment; such a renormlization is still within the uncertainties of the flux determination by MB. Applying the same correction to the pion data brings them into better overall agreement with our theoretical results  \cite{Lalakulich:2012hs}. However, even then there is still an indication of the difference in shape of the experimental and theoretical distributions. This has to be clarified through a new look at the data. The special shape of the distribution obtained in the calculations is a consequence of FSI and these final state interactions should be the same as in photoproduction of pions. Upcoming new data from electroproduction at JLAB could also be most useful to clarify this question.

\section{Interactions in the SIS regime}
The long baseline experiments MINOS and NO$\nu$A at Fermilab both work at a higher neutrino energy than MiniBooNE. While NO$\nu$A uses a relatively sharp flux peaking at about 2 GeV, MINOS works with a broader distribution of incoming neutrino energies with a peak around 3 GeV and an average around 5 GeV. Fig.~\ref{fig:compare-origin-MINOS-MB-3} illustrates that while NO$\nu$A has about equal parts of QE scattering, pion production and DIS the MINOS experiment is clearly dominated by DIS, however there is still a significant amount of pion production and a small contribution of QE scattering. Both experiments work in the transition area between resonance dominated physics and DIS that is located above a baryon mass of about 2 GeV. While concepts of perturbative QCD work very well in the asymptotic (Bjorken) limit \cite{Conrad:1997ne}, this so-called shallow-inelastic scattering (SIS) is a much more complicated regime. Here, for example, 2-meson channels, for which usually there are only very few data available, may become important. This problem of missing elementary input then affects all predictions for neutrino-nucleus interactions in this region.

\begin{figure}[htb]
\includegraphics[width=\columnwidth]{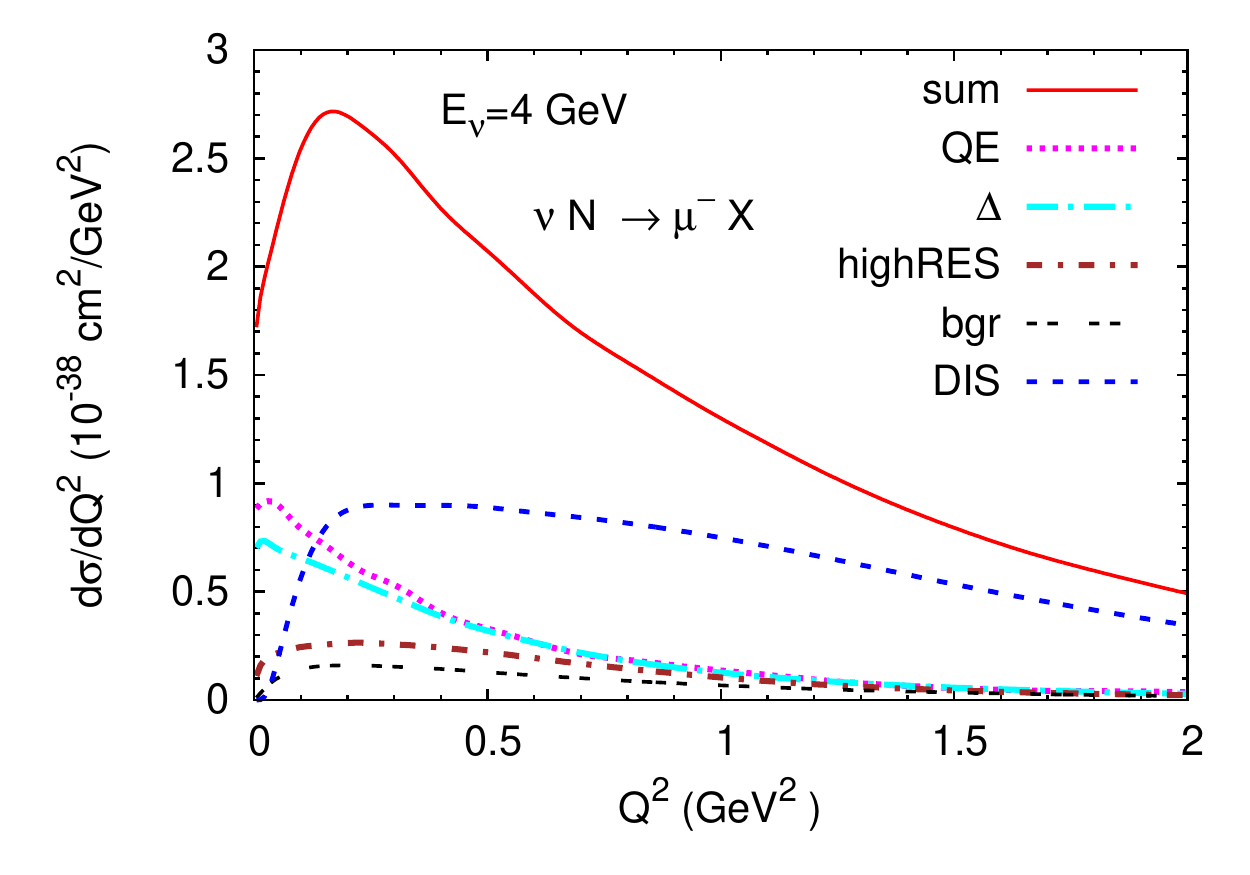}
\caption{(Color online) Cross section $\dd \sigma/\dd Q^2$ per nucleon for CC neutrino scattering off an isoscalar target
 for $E_{\nu}=4\GeV$.}
\label{fig:isoscalar-dsidQ2}
\end{figure}
Fig.\ \ref{fig:isoscalar-dsidQ2} shows the $Q^2$ dependence of the various contributions to the cross section for an incoming energy of 4 GeV. One sees that even at this higher energy the $\Delta$ and QE scattering contribute to the cross section, but they are dominant only at small $Q^2 < 0.4 \GeV^2$. For larger $Q^2$ DIS clearly dominates. This different behavior may offer a possibility to distinguish between the other reaction mechanisms and DIS if the $Q^2$ reconstruction can be done reliably.

\begin{figure}[h!bt]
\includegraphics[width=\columnwidth]{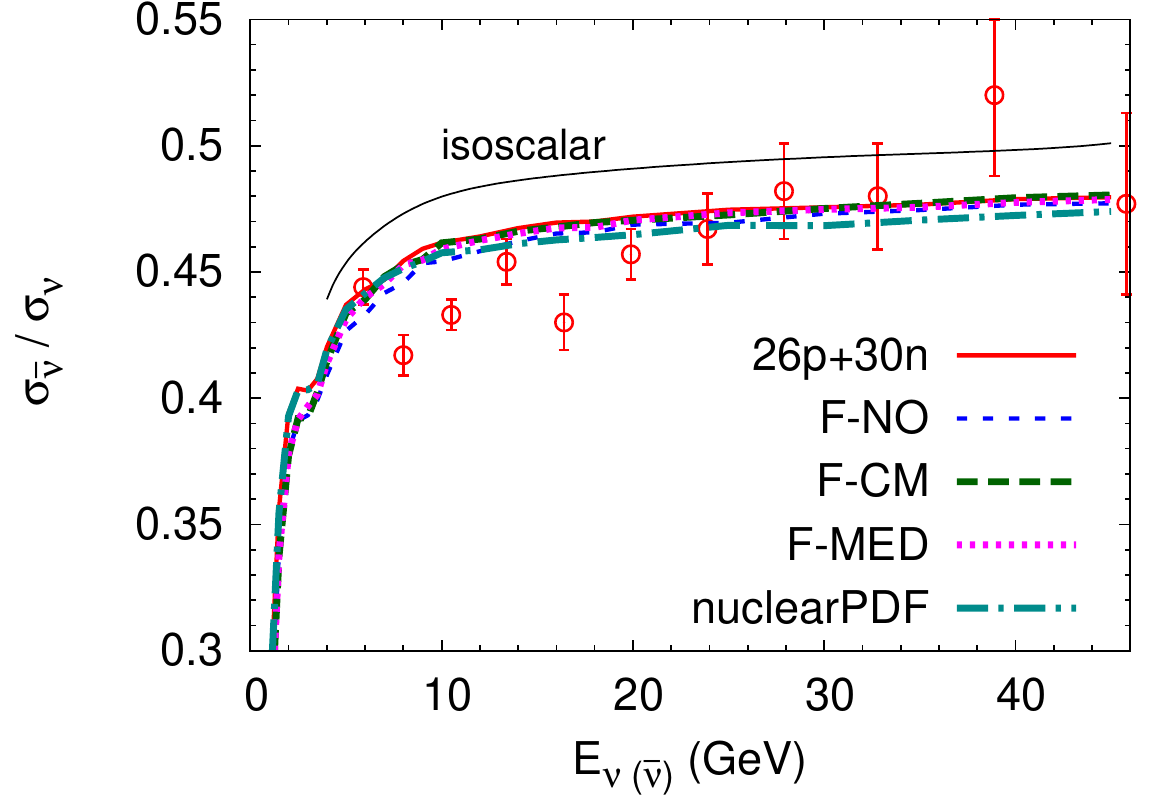}
\caption{(Color online) Ratio of the antineutrino to neutrino cross sections in scattering off iron. Also shown is the calculated ratio for an isoscalar target.
Experimental data are taken from \protect\cite{Adamson:2009ju} (MINOS, open circles). Figure taken from \protect\cite{Lalakulich:2012gm}.}
\label{fig:compare-nuclear-nu-barnu-ratio}
\end{figure}
At high energies, in the asymptotic QCD regime, the ratio of antineutrino to neutrino cross sections approach the constant value $\approx 0.5$ at energies above about 25 GeV \cite{Conrad:1997ne}. Theory and data show that this is not the case in the SIS region. Fig.\ \ref{fig:compare-nuclear-nu-barnu-ratio} shows that this ratio increases steeply as a function of neutrino energy for energies below about 10 GeV. This steep rise is due to the fact that for an antineutrino both the QE and the $\Delta$ resonance excitation cross sections are significantly lower than for a neutrino, leading to a small ratio. With increasing neutrino energy these components die out and DIS takes over a dominating role. For DIS the ratio then is given (for an isoscalar target) by the asymptotic value of about 0.5. The various curves in the figure illustrate different treatments of in-medium effects; that they all agree with each other shows that this ratio is fairly independent of nuclear effects.
\begin{figure}[h!bt]
\centering
\hfill
\includegraphics[width=0.6\textwidth]{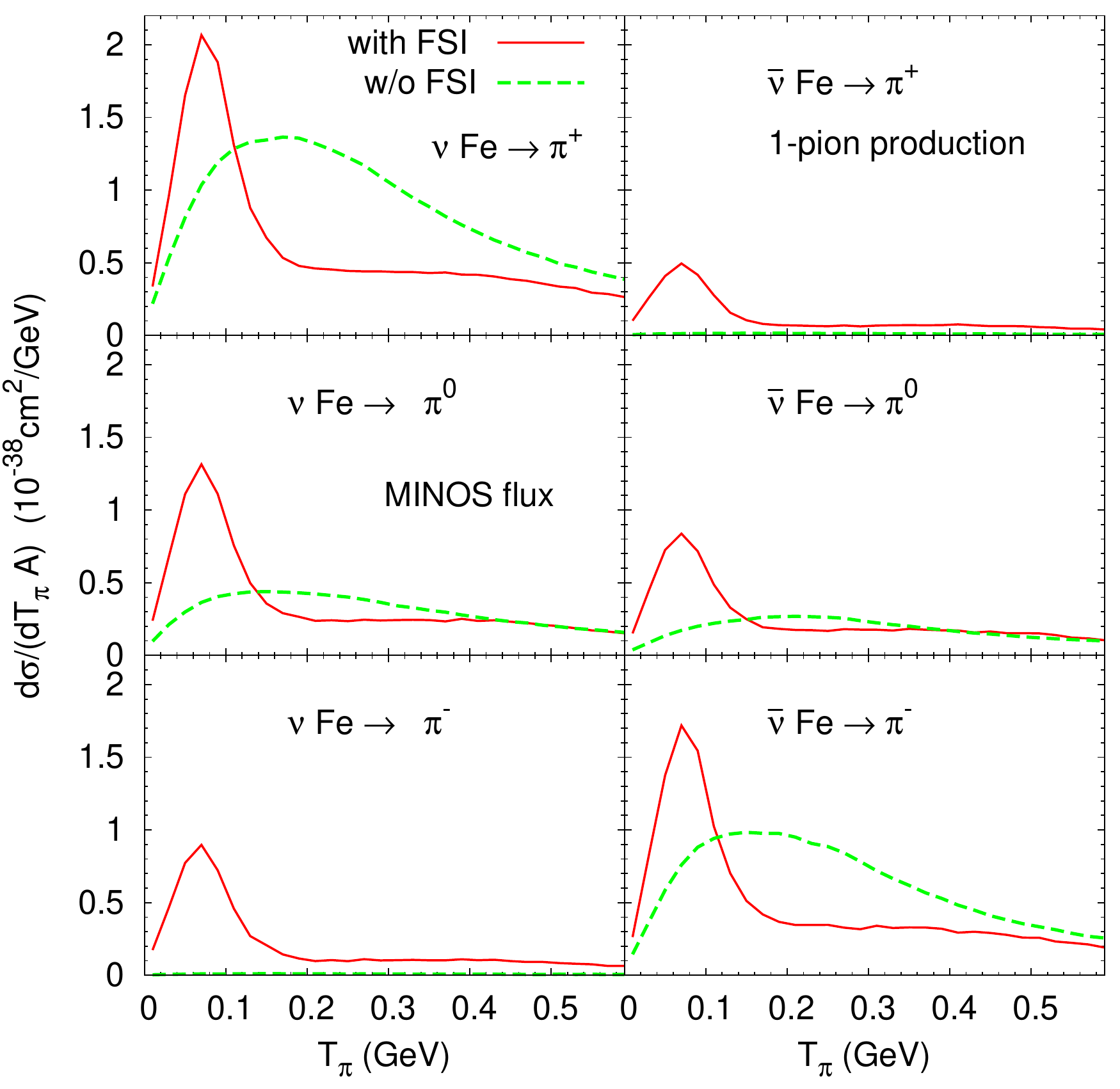}
\hfill
\caption{(Color online) Pion kinetic energy distributions per target nucleon
for neutrino and antineutrino scattering off iron for \onepion production
(one pion of the indicated charge and no other pions are produced). Figure taken from \cite{Lalakulich:2012gm}.}
\label{fig:MINOS-ekin-with-wo-FSI-1-pion}
\end{figure}

\begin{figure}[h!bt]
\centering
\hfill
\includegraphics[width=0.6\textwidth]{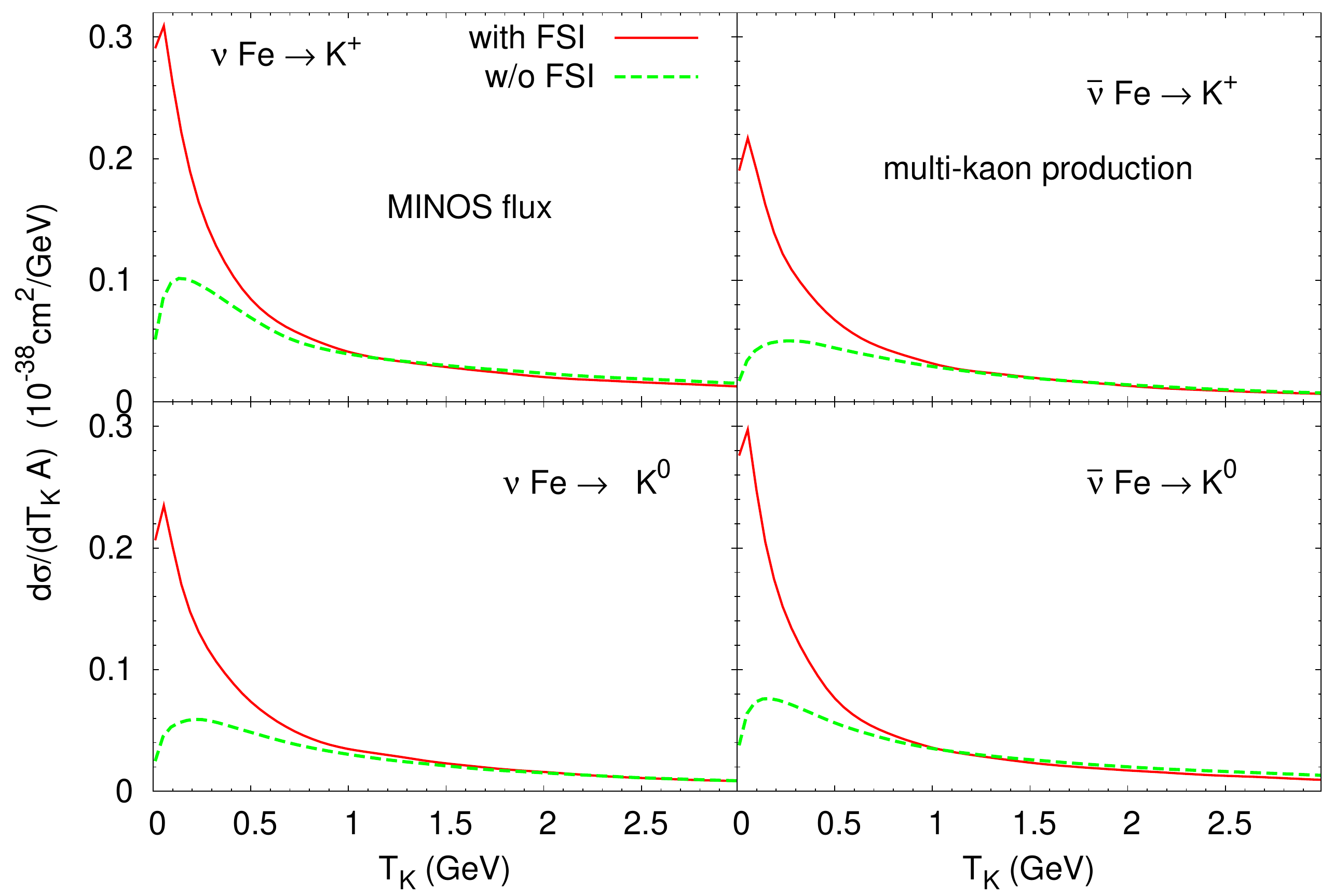}
\hfill*
\caption{(Color online) Kinetic energy distributions per target nucleon for multi-kaon
(at least one kaon of a given charge and any other hadrons) production
in neutrino and antineutrino scattering off iron. Figure taken from \protect\cite{Lalakulich:2012gm}.}
\label{fig:MINOS-ekin-with-wo-FSI-kaon-MULTI}
\end{figure}

In Ref.\ \cite{Lalakulich:2012gm} we have given detailed predictions for spectra of knock-out nucleons and produced mesons for both the MINOS and the NO$\nu$A fluxes. Here, as an example, we just show the distributions for pions and for kaons in Figs.\ \ref{fig:MINOS-ekin-with-wo-FSI-1-pion} and \ref{fig:MINOS-ekin-with-wo-FSI-kaon-MULTI}, resp.,  for the MINOS flux.

The pion spectra resemble those discussed earlier for the MB flux. They all exhibit the suppression around 0.2 GeV due to pion FSI. This just shows that even though these pions have originally been produced by quite a different mechanism as compared to the situation at MB (DIS vs. $\Delta$ resonance) the final state interactions for pions are so strong that they essentially wipe out any memory of the production process. In this situation the observable spectra are determined only by the FSI. As discussed in \cite{Lalakulich:2012gm} very few of the pions finally observed are actually the same as those produced in the primary reaction process. Instead, pion absorption and reemission takes place and helps again to shadow the elementary production process.

This is even more so for kaons. While there are new calculations of neutrino-induced exclusive kaon production cross sections on the nucleon \cite{RafiAlam:2010kf} these elementary processes will be very hard -- if not impossible - to observe. Since the neutrino flux always involves broad distributions with tails towards higher energies the observed kaon production rate is always dominated by DIS followed by strong secondary interactions. This is again discussed in some detail in \cite{Lalakulich:2012gm}. Here we just show the spectra in Fig.\ \ref{fig:MINOS-ekin-with-wo-FSI-kaon-MULTI}. All these spectra show a pile-up at low kinetic energies due to multiple FSI. Note that the results shown here are for $K^+$ and $K^0$. For both, because of strangeness conservation, the absorption on a nucleon in the nuclear target is very small, but they can easily undergo charge transfer processes $K^+ n \to K^0 p$ or inelastic reactions such as $K^+  N \to K^+ \Delta$ which leads to a loss of kaon energy and simultaneous pion production.

\section{Extraction of oscillation parameters}
\begin{figure}[!hbt]
\includegraphics[width=\columnwidth]{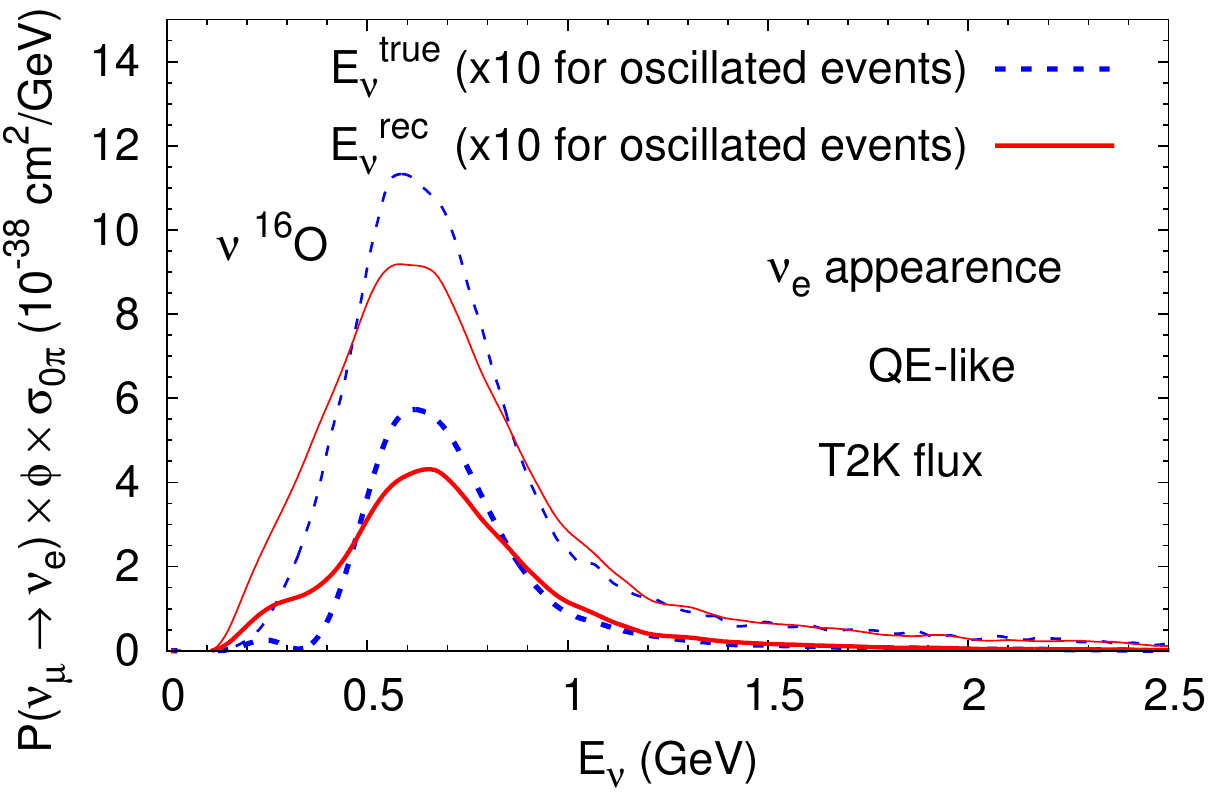}
\caption{(Color online) The event rate, i.e.\ flux times cross section, for electron neutrino appearance measurements. The oscillated event curves have been multiplied by a factor of 10 to enhance the visibility of their difference. Figure taken from \protect\cite{Lalakulich:2012hs}.}
\label{fig:T2Kflux-oscillations-Che-before-after-mue}
\hfill
\end{figure}
\begin{figure}[!hbt]
\includegraphics[width=\columnwidth]{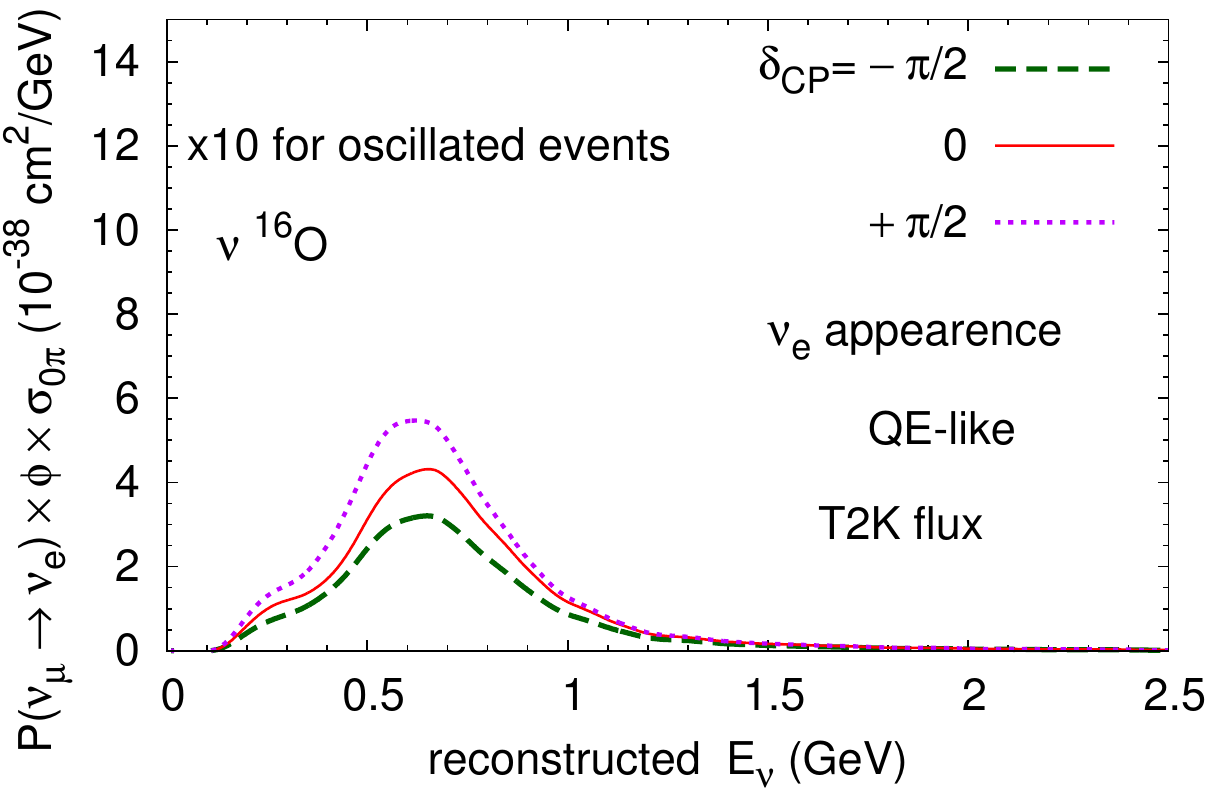}
\caption{(Color online) Same as Fig.\ \ref{fig:T2Kflux-oscillations-Che-before-after-mue} for various CP violation phases.  
Figure taken from \protect\cite{Lalakulich:2012hs}.}
\label{fig:T2Kflux-oscillations-Che-mue-deltaCP}
\end{figure}
After having observed this rather significant influence of the energy reconstruction on the observed cross sections the question immediately arises how this affects the extraction of neutrino oscillation properties. This is indicated in Fig.\ \ref{fig:T2Kflux-oscillations-Che-before-after-mue} which shows the calculated event rate for electron appearance in the T2K experiment which uses QE-like events as signal. Again, it can be seen that the event rate as a function of reconstructed energy has more strength at lower energies and less in the peak region than that as a function of true energy.
How this extra strength affects the actual oscillation parameters has not been explored yet, but there are first attempts by D. Meloni et al.\ \cite{Meloni:2012fq} to investigate this question. Here it is sufficient to note that the difference between the true-energy and the reconstructed-energy results is similar to that expected from varying the phase $\delta_{\rm CP}$ that controls the possible $CP$ violation in electroweak interactions (see Fig.\ \ref{fig:T2Kflux-oscillations-Che-mue-deltaCP}). Any experiment aiming for a determination of $\delta_{\rm CP}$ with neutrinos alone would thus have a hard time to achieve the necessary sensitivity to that phase.

\section{Summary}
Neutrino-nucleus interactions in the energy range of the MiniBooNE, T2K, NO$\nu$A, MINER$\nu$A and MINOS experiments are sensitive to various elementary interactions. QE scattering, two-body interactions connected with meson exchange currents, pion production through nucleon resonances (including background) and deep inelastic scattering all play a role. A theoretical treatment thus has to be able to deal with all of them with equal accuracy. This is so important because neutrino experiments in the past tended to subtract some of these processes as background from their data. The remaining 'data' were then already influenced by possible shortcomings of the generators.

We are now in a position to deal with the actual data, e.g.\ directly the QE-like cross sections measured by MiniBooNE, and not just the extracted 'data'. For the MiniBooNE this requires primarily a good treatment of QE scattering \emph{and} pion production, both from resonances, background and (to a minor extent) from DIS. In this talk we have discussed the theoretical results and compared them with the MB pion data. A significant sensitivity to the elementary input is found. If many-body contributions to pion production are not large then only the old BNL elementary pion production data seem to be compatible with the MB data. An ultimate clarification of this question can, however, come only from new data for neutrino interactions with single nucleons.

At the higher energies so-called shallow inelastic processes, for which very few elementary data are known, complicate the picture. QE-scattering and resonance excitations still contribute even at a few GeV neutrino energy to the total cross section. However, these events are localized at lower momentum transfers whereas at the higher $Q^2$ DIS takes over. Particle production in this energy regime is sensitive not only to these primary production processes, but also to strong FSI. This is true not only for the strongly interacting pions, but also for positively charged and neutral kaons which are (erroneously) often believed to suffer very few FSI because of strangeness conservation.  As a consequence, one has to conclude that any measurements of pions or kaons in the MINER$\nu$A or NO$\nu$A experiments do not yield information on the elementary cross section.

Finally we have illustrated how shortcomings in the identification of QE scattering and the energy reconstruction that is built on it can affect the oscillation parameters. These uncertainties clearly affect the extraction of the $CP$-violating phase.

This work was supported by DFG and by BMBF.

\bibliographystyle{aipproc}
\bibliography{nuclear}
\end{document}